\documentclass[aps, prb, superscriptaddress, twocolumn, 10pt, amsfonts, amsmath, amssymb,
 reprint,
]{revtex4-2}

\usepackage{graphicx}
\usepackage{dcolumn}
\usepackage{bm}
\usepackage{hyperref}

\usepackage{lipsum}
\usepackage{braket}
\usepackage{xcolor}

\begin{document}

\preprint{APS/123-QED}

\title{Transport-based fusion that distinguishes between \\ Majorana and Andreev bound states}

\author{Maximilian Nitsch}
\affiliation{Division of Solid State Physics and NanoLund, Lund University, S-22100 Lund, Sweden}
\author{Rubén Seoane Souto}
\affiliation{Division of Solid State Physics and NanoLund, Lund University, S-22100 Lund, Sweden}
\affiliation{Center for Quantum Devices, Niels Bohr Institute, University of Copenhagen, DK-2100 Copenhagen, Denmark}
\affiliation{Instituto de Ciencia de Materiales de Madrid (ICMM), Consejo Superior de Investigaciones Científicas (CSIC),
Sor Juana Inés de la Cruz 3, 28049 Madrid, Spain.}
\author{Stephanie Matern}
\affiliation{Division of Solid State Physics and NanoLund, Lund University, S-22100 Lund, Sweden}
\author{Martin Leijnse}
\affiliation{Division of Solid State Physics and NanoLund, Lund University, S-22100 Lund, Sweden}
\affiliation{Center for Quantum Devices, Niels Bohr Institute, University of Copenhagen, DK-2100 Copenhagen, Denmark}

\date{\today}

\begin{abstract}
It has proven difficult to distinguish between topological Majorana bound states and nontopological Andreev bound states and to measure the unique properties of the former. In this work, we aim to alleviate this problem by proposing and theoretically analyzing a new type of fusion protocol based on transport measurements in a Majorana box coupled to normal leads. The protocol is based on switching between different nanowire pairs being tunnel coupled to one of the leads. For a Majorana system, this leads to switching between different states associated with parity blockade. The charge being transmitted at each switch provides a measurement of the Majorana fusion rules. Importantly, the result is different for a system with nontopological Andreev bound states. The proposed protocol only requires measuring a DC current combined with fast gate-control of the tunnel couplings. 
\end{abstract}

\maketitle


\section{Introduction}

Finding Majorana bound states (MBSs) in topological superconducting systems has been an intensely pursued goal in condensed matter physics for over a decade \cite{Kitaev_2001, NayakReview, Wilczek_nature2009, Alicea_RPP2012, Leijnse_Review2012, BeenakkerReview_20}. A promising platform to find MBSs is based on one-dimensional superconductor-semiconductor hybrid structures \cite{Oreg_PRL2010, Lutchyn_PRL2010, Lutchyn_NatRev2018, flensberg2021engineered}. There have been encouraging results in the form of observations of zero-bias peaks consistent with Majorana physics, see, e.g. Refs. \cite{Mourik_science2012, deng2012anomalous, finck2013anomalous, deng2016majorana, Nichele_PRL2017, Lutchyn_NatRev2018, Vaitiekenas_Science2020, Yazdani_Science2023}. However, these experiments offer no definite proof of topological MBSs, as topologically trivial systems hosting Andreev bound states (ABSs) can exhibit similar features \cite{Prada_PRB2012, Kells_PRB12, Moore_PRB18,reeg2018zero, Awoga_PRL2019, Vuik_SciPost19, Pan_PRR20, Prada_review,hess2021local}.

To obtain definite proof for the topological nature of the observed states it is necessary to probe their nonabelian properties. Despite a large number of theoretical proposals, see Refs. \cite{Bonderson_PRL2008,Alicea_NatPhys2011, Clarke_PRB2011, Sau_PRB2011, Flensberg_PRL2011, vanHeck_NJP2012, Aasen_PRX2016, Vijay_PRB2016, Karzig_PRB2017, Hell_PRB2017, Stern_PRL2019, Clarke_PRB2017} for a few examples, there has been no experimental realization of a braiding protocol. Instead of aiming for braiding, a simpler but conceptually related concept is fusion for which there also exist various suggestions for experimental realizations \cite{Alicea_NatPhys2011, Aasen_PRX2016, Clarke_PRB2017, Beenaker_SciPost2019, Souto_SciPost2022, Liu_PRB2023}. The idea of fusion protocols is to measure the MBS system in different basis combinations, thereby effectively accessing the nonabelian properties. Unfortunately, some fusion protocols can give the same outcome for zero-energy ABSs as for MBSs \cite{Clarke_PRB2017, tsintzis2023roadmap} and therefore do not offer sufficient proof of a nonabelian topological phase.

The main goal of this paper is to provide a proposal for a Majorana fusion experiment that explicitly distinguishes between topological and trivial systems. Our proposal is based on DC transport measurement and does not require fast or single-shot read-out, only fast gate voltage pulses.

The platform we consider is a Majorana box qubit \cite{Beri_PRL2012, Plugge_NJP2017} a candidate for scaleable topologically protected quantum computing \cite{Vijay_PRX2015, Landau_PRL2016, Plugge_PRB2016, Karzig_PRB2017}. We connect it in a transport setup with two normal metallic leads (source, drain) and tune the connections between the source and box via a magnetic flux to establish parity blockade. Parity blockade, destructive interference between two paths via two Majoranas, was introduced in previous works on Majorana box qubits connected to quantum dots \cite{munk2020parity, steiner2020readout, Schulenborg_PRB2021, Schulenborg_PRB2023} and in transport setups \cite{Nitsch_PRB2022}.
\begin{figure}[t]
\includegraphics[width=0.9\columnwidth]{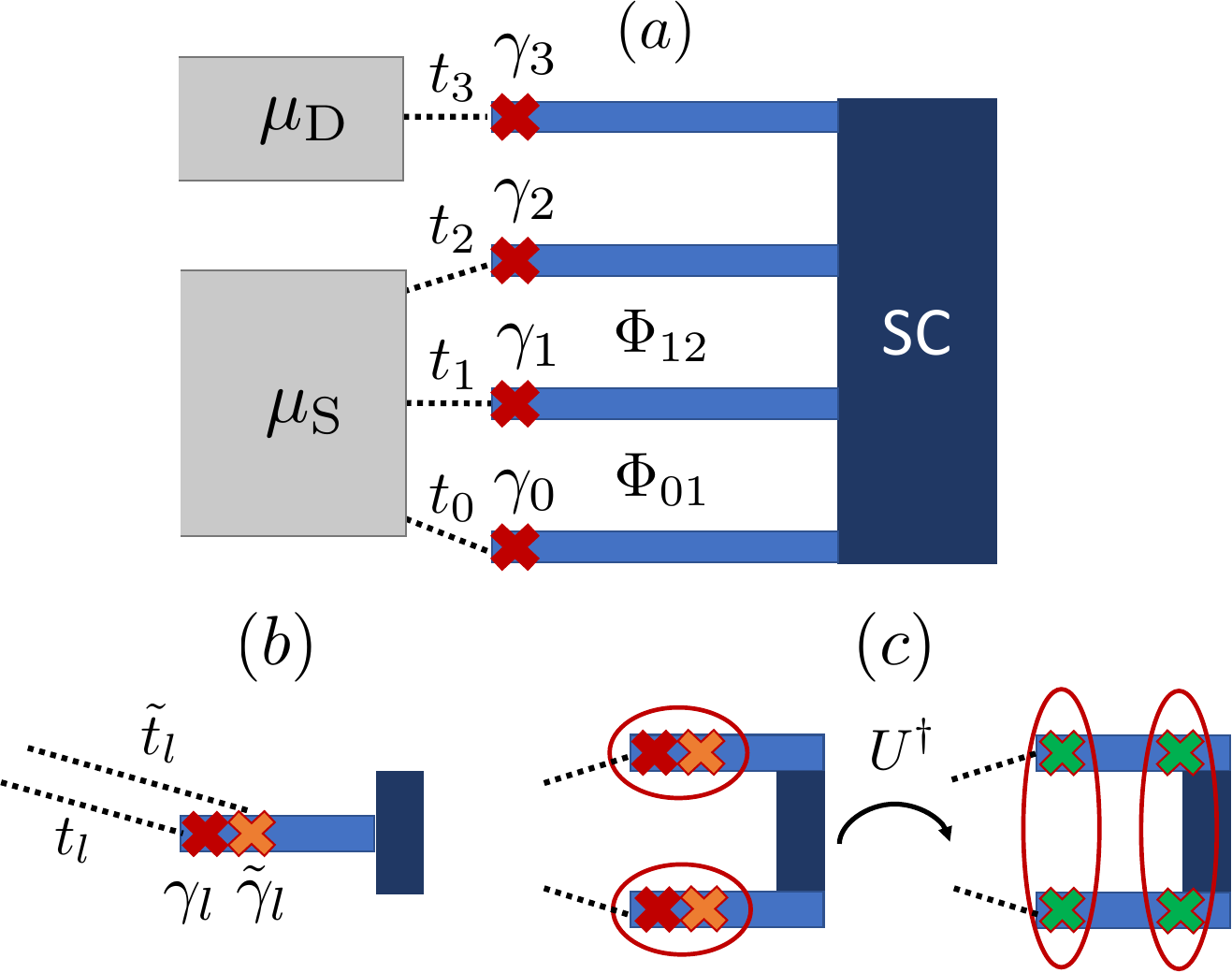}
\caption{$(a)$ Floating Majorana box with four topological nanowires hosting three MBSs $\gamma_{0,1,2}$ connected to the source (S) and one $\gamma_3$ to the drain (D) via tunnel couplings $t_{0,1,2,3}$. Two magnetic fluxes $\Phi_{01}, \Phi_{12}$ are threaded through the loops $t_0, t_1$ and $t_1, t_2$ to adjust the relative phases. $(b)$ Tunnel couplings $t_l, \tilde{t}_l$ to two Majorana operators $\gamma_l, \tilde{\gamma}_l$ describing an ABS in a topologically trivial system. $(c)$ Basis rotation $U^\dagger$ to describe two ABSs via two coupled and two decoupled MBSs.}
\label{fig:figure1}
\end{figure}

The proposed fusion protocol works as follows: In a transport setup with three MBS wires coupled to the same lead, parity blockade projects the box state onto well-defined blocking states. By repeatedly switching between different configurations of the lead-wire couplings, the system alternates between different blocking states. The projection of one blocking state onto another is similar to MBS fusion and, at the same time, determines the probability for a single electron to be transferred. A DC current measurement reveals the average outcome of the fusion protocol.

The paper starts in Sec.~\ref{sec:SetupAndTransport} with an introduction to the system and transport setup, the concept of parity blockade and the quantum master equation (QME) used for the transport simulations. Afterwards, we investigate how to employ parity blockade to distinguish between topological and trivial systems in Secs. \ref{sec:StationaryProtocols} and \ref{sec:fusion}. First, we show that the possibility to block the current via parity blockade is not sufficient to distinguish topological from trivial systems (Sec.~\ref{subsec:BlockadeManifold}). In order to distinguish these cases, we first introduce a simple protocol based on establishing parity blockade followed by switching off different tunnel couplings (Sec.~\ref{subsec:currentDistinction}), before moving on to the fusion protocol (Sec.~\ref{sec:fusion}). 

\section{Setup and transport description}
\label{sec:SetupAndTransport}

\subsection{Majorana box}
\label{sec:MBSbox}
We consider a Majorana box consisting of four topological nanowires hosting MBSs at their ends, connected by a small piece of superconductor in the trivial regime, see Fig.~\ref{fig:figure1}$(a)$. The entire system, wires and connecting superconductor, is floating and has a charging energy $E_C$, which we take to be the largest energy scale of the problem. A gate voltage $V_g$ capacitively coupled to the box tunes the amount of energetically favorable charge $n_g$, thereby setting the electron number $N$ of the ground state. Furthermore, we include small but finite energy splitting in the degenerate ground state sector due to exponentially small overlaps $\varepsilon_{lk}$ between MBSs $\gamma_l, \gamma_{l^\prime}$. The Majorana box Hamiltonian reads
\begin{align}\label{eq:HMB}
    H_\text{MB}=\frac{i}{2} \sum_{l l^\prime} \varepsilon_{ll^\prime} \gamma_{l}\gamma_{l^\prime} + E_C(N-n_g)^2,
\end{align}
where the sum runs over $l < l^\prime$. We note that there will in general be four additional MBSs at the points where the nanowires meet the trivial superconductor, and the overlaps with these MBSs might be larger than between MBSs at different nanowire ends. However, in the limit where the overlaps set the smallest energy scale of the problem, the  results presented below remain qualitatively the same independent of which MBSs couple, and these additional MBSs can safely be neglected~\cite{comment1}.

To enable charge flow, we connect the box to source (S) and drain (D) leads, described by noninteracting electrons, $H_{res} =\sum_{r=S,D} H_r$, with $H_r = \sum_{k} \xi_{rk} c_{rk}^\dagger c_{rk}$. The operators $c_{rk}^\dagger, c_{rk}$ create/destroy an electron in lead $r$ with momentum $k$ and energy $\xi_{rk}$. We neglect spin, assuming that either the leads are spin polarized by a large magnetic field or all MBSs have equal spin polarization~\cite{Nitsch_PRB2022}. The leads are characterized by a temperature $T_r = T$ and a symmetrically applied bias voltage setting the chemical potentials $\mu_{S,D} = \pm V_b/2$.

The leads are connected to the box via tunnel amplitudes $t_{lr}$ between the $l$th MBS and lead $r$
\begin{align}\label{eq:HMBT}
    H_\text{MB}^T = \sum_{lrk}\gamma_{l}\left( t_{lr}c_{rk} - t_{lr}^* c_{rk}^\dagger \right).
\end{align}
We consider wide-band leads with energy-independent tunnel couplings. Only one MBS ($\gamma_3$) is coupled to the drain, while three MBSs ($\gamma_{0, 1, 2}$) are coupled to the source, and to enable parity blockade these MBSs need to connect to the same channel of the source~\cite{Nitsch_PRB2022}. The relative phases of $t_l$ and $t_{l^\prime}$ can be controlled via fluxes $\Phi_{ll^\prime}$.

Unless stated otherwise, we assume $|t_l| = t > 0$ for all $l$. This introduces a tunneling rate $\Gamma = 2 \pi \nu \, t^2$, where $\nu$ is the density of states in the leads, which we take to be energy-independent and equal for source and drain.

\subsection{Andreev box\label{sec:ABSbox}}
For comparison we will consider also an "Andreev box", i.e., a system that is equivalent to the Majorana box, but where the (near) zero-energy states are nontopological ABSs. To facilitate the comparison to the Majorana box, we decompose each ABS into two MBSs, $\gamma_l$ and $\tilde{\gamma}_l$, see Fig.~\ref{fig:figure1}$(b)$. This allows writing the Hamiltonian for the Andreev box in a very similar way to Eq.~(\ref{eq:HMB})
\begin{align}\label{eq:HAB}
    H_\text{AB}=\frac{i}{2}\sum_l \varepsilon_l \gamma_{l}\tilde{\gamma}_{l} + E_C(N-n_g)^2,
\end{align} 
where the energy $\varepsilon_l$ of ABS $l$ is included as an overlap between the constituent MBSs and we have neglected overlaps between different ABSs. We assume that $\varepsilon_l$ constitutes the smallest energies in the problem, which is the case where ABSs and MBSs are hard to distinguish.

The source and drain leads are described in exactly the same way as for the Majorana box and their coupling to the Andreev box is given by
\begin{align}\label{eq:HABT}
    H_\text{AB}^T = \sum_{lrk}\left[\gamma_{l}\left( t_{lr}c_{rk} - t_{lr}^* c_{rk}^\dagger \right) + 
    \tilde{\gamma}_{l}\left( \tilde{t}_{lr}c_{rk} - \tilde{t}_{lr}^* c_{rk}^\dagger \right)\right],
\end{align}
where for ABS $l$, electrons can tunnel into/out of both MBS constituents $\gamma_l$ and $\tilde{\gamma}_l$ with amplitudes $t_l$ and $\tilde{t}_l$, see Fig.~\ref{fig:figure1}$(b)$. For nontopological ABSs, even if $\varepsilon_l$ is small, $\gamma_l$ and $\tilde{\gamma}_l$ have similar spatial distribution~\cite{Penaranda_PRB2018, Avila_CommPhy2019, Prada_Nature2020} and one expects both $t_l$ and $\tilde{t}_l$ to be finite. The situation is complicated by the fact that both the relative amplitudes and phases of $t_l$ and $\tilde{t}_l$ are important, but there is freedom in choosing the way each ABS is decomposed into MBSs which allows us to fix one of them. We choose the Majorana basis such that $|t_l| = |\tilde{t}_l|$, leaving the relative phase
\begin{equation}
\label{eq:theta}
    \tilde{t}_l = e^{i \theta_l} \, t_l,
\end{equation}
as a parameter that is determined by the specific realization of the system. We will show in Secs. \ref{sec:StationaryProtocols}, \ref{sec:fusion} that the value of $\theta_l$ quantifies how well an ABS resembles a single localized MBS.

\subsection{Transport description via QME}
To calculate the transport properties of the Majorana and Andreev box, we use a QME approach based on leading-order perturbation theory in $\Gamma$. After tracing out the lead degrees of freedom, the equation of motion of the Majorana box density matrix is given by
\begin{equation}
    \label{eq:QME}
    \partial_t \rho = \mathcal{L} \rho = -i [H_{\text{XB}}, \rho ] + \mathcal{D} \, \rho.
\end{equation}
The Liouvillian $\mathcal{L}$ consists of two general parts, the unitary time evolution, determined by the box Hamiltonian $H_{\text{XB}}$, either $H_\text{MB}$ (Eq.~\ref{eq:HMB}) or $H_\text{AB}$ (Eq.~\ref{eq:HAB}), and a dissipative part $\mathcal{D}$ introduced by the coupling to the leads.

The QME we are using is a Redfield-type approach (called 1st order von Neumann in Ref.~\cite{Kirsanskas_CPC2017}), which is equivalent to the first order of real-time diagrammatics \cite{konig1997cotunneling, leijnse2008kinetic, schoeller2009perturbative}. The procedure for the numerical solution is as follows: We start by diagonalizing $H_\text{XB}$, thereby obtaining the eigenenergies and many-body eigenstates of the disconnected system. In the next steps, we express Eq.~(\ref{eq:QME}) in the many-body eigenbasis. Based on the tunneling Hamiltonian in Eqs.~(\ref{eq:HMBT}) and (\ref{eq:HABT}) we calculate the tunnel matrix elements between eigenstates $\ket{a}$ and $\ket{b}$ as
\begin{align}
    \label{eq:T_r}
T_{b \rightarrow a }^{r, \text{MB}} = \sum_{l} t_{lr} \braket{a | \gamma_{l} |b}
\end{align}
for the Majorana box and as
\begin{align}
    \label{eq:T_r_Andreev}
T_{b \rightarrow a}^{r, \text{AB} } = \sum_{l} \left[ t_{lr} \braket{a | \gamma_{l} |b} + \tilde{t}_{lr} \braket{a | \tilde{\gamma}_{l} |b} \right]
\end{align}
for the Andreev box. Afterwards follows the calculation of the dissipative part $\mathcal{D}$ of Eq.~(\ref{eq:QME}), according to the 1st order von Neumann method. See Appendix~\ref{sec:appendix1vN} and \cite{Kirsanskas_CPC2017} for details on the calculation of $\mathcal{D}$. Equation (\ref{eq:QME}) is now expressed in superoperator notation
\begin{equation}
    \partial_t |\rho ) = \hat{\mathcal{L} } \, |\rho ).
\end{equation}
The density matrix is rearranged into the vector $| \rho )$ and the Liouvillian is expressed as a matrix $\hat{\mathcal{L}}$ called the kernel.

We obtain the solution of this equation via numerical diagonalization of the kernel. As the kernel is non-Hermitian, the left and right eigenvectors $|l_h)$, $|r_h)$ for a given eigenvalue $\chi_h$ are not guaranteed to be the same. We calculate both and obtain the solution as
\begin{equation}
    \label{eq:SolutionQME}
    |\rho)(t) = |\rho)_\text{ss} +  \sum_{h>0} e^{\chi_h t} c_h |r_h),
\end{equation}
where $c_h$ is obtained from the initial state $|\rho_0 )$ as
\begin{equation}
    \label{eq:IntialStateWeights}
    c_h = (l_h | \rho_0).
\end{equation}
The solution consists of two parts: The stationary state solution $|\rho)_\text{ss}$ and the finite-time contribution. Due to the non-Hermiticity of the kernel, its eigenvalues are complex-valued. In order for a physical solution, there are two conditions that have to be fulfilled by the eigenvalues. First, there needs to be a zero eigenvalue $\chi_0 = 0$ yielding the stationary state solution. Second, the real-valued parts of the remaining eigenvalues are strictly negative and lead to a decay of all finite-eigenvalue contributions. Depending on the system, this decay can be decorated with oscillations due to an imaginary part. As a last step, we calculate the current through the system from the density matrix as in \cite{Kirsanskas_CPC2017}.

\subsection{Parity blockade}
\label{subsec:parityBlockade}
As in previous studies on Majorana box qubits connected to quantum dots \cite{munk2020parity, steiner2020readout, Schulenborg_PRB2021, Schulenborg_PRB2023} and in transport setups \cite{Nitsch_PRB2022}, we use the magnetic flux $\Phi$ as a tunable parameter to establish the parity blockade. In the most simple case of the Majorana box, 
the source connects to the MBSs $\gamma_0$ and $\gamma_1$ with the same strength for the tunnel couplings $|t_0| = |t_1| = t$ and is disconnected from $\gamma_2$. We describe the systems via the fermionic occupation $n_{01} = f_{01}^\dagger f_{01}$ with $f_{01} = \gamma_0 + i \gamma_1$. The combined tunnel matrix element for an electron to enter the system via $f_{01}, f_{01}^\dagger$ reads
\begin{align}
    T_{0 \rightarrow 1}^\text{S,MB} = t(1 + i e^{i \phi}), && T_{1 \rightarrow 0}^\text{S,MB} = t(1 - i e^{i \phi}),
\end{align}
where $\phi$ is tuned via a magnetic flux $\Phi_{01}$ threaded between the connections from the source to $\gamma_0$ and $\gamma_1$. We use it to establish constructive or destructive interference between the two available paths. For example, tuning the phases to $\phi = \frac{\pi}{2}$ results in $T_{0 \rightarrow 1}^\text{S,MB} = 0$ and therefore prohibits the transition $n= 0 \rightarrow n = 1$, which we refer to as the establishing of parity blockade.

Returning to the full Majorana box (Sec.~\ref{sec:MBSbox}) we choose the fermionic basis by combining MBSs $\gamma_0$ with $\gamma_1$ and $\gamma_2$ with $\gamma_3$, defining the Fock states $\ket{n_{01} n_{23}}$. Parity blockade at the source projects the system on a state with total even parity ($-1^{n_{01} + n_{23}} = 1$) spanned by $\ket{00}$ and $\ket{11}$. The exact form of the blocking state depends on the way it is established via choices of $t_0, t_1, t_2$.

In the Andreev box (Sec.~\ref{sec:ABSbox}) the intuitive way to understand parity blockade for zero-energy ABSs is that there always exists a unitary rotation of the Majorana basis to effectively only couple two MBSs to the lead \cite{Schulenborg_PRB2023}, see Fig.~\ref{fig:figure1}$(c)$. In our case, the ABSs and therefore also the effective MBSs are on separate sites with a magnetic flux threaded in between, such that we can use the connections and flux to establish parity blockade.

\section{Stationary state protocols}
\label{sec:StationaryProtocols}
In this section, we focus on results obtained by solving for the stationary state current. In Sec.~\ref{subsec:BlockadeManifold} we conclude that observing parity blockade is insufficient to distinguish the Majorana box from the Andreev box. Afterwards, we provide a protocol allowing their distinction in Sec.~\ref{subsec:currentDistinction}.

In the remainder of the paper, the system parameters are chosen as follows. First of all, the temperature of the leads is assumed to be far larger than the tunneling rate to the leads $T = 10^2 \, \Gamma$. We set the chemical potential $\mu_{S/D} = \pm 10^3 \, \Gamma$. All energies resulting from MBS overlaps are of the order of $\varepsilon = 10^{-3} \, \Gamma$. The box is tuned via electrostatic gates to a degeneracy point $n_g = N + \frac{1}{2}$ between $N$ and $N+1$ charges. The exact values for the gate and bias voltage don't influence the general transport behavior. But to enable transport, we must make sure that the system is in the conducting regime and not Coulomb blockaded.

For the Majorana box, the overlaps are zero except for the combinations $\varepsilon_{01} = 1.0 \, \varepsilon$, $\varepsilon_{12} = 1.5 \, \varepsilon$, $\varepsilon_{23} = 2.0 \, \varepsilon$. In the Andreev box, we choose the overlaps of MBSs on each site as $\varepsilon_{0} = 0.5 \, \varepsilon$, $\varepsilon_{1} = 1.0 \, \varepsilon$, $\varepsilon_{2} = 1.5 \, \varepsilon$, $\varepsilon_{3} = 2.0 \, \varepsilon$. The exact choice does not influence the general behavior, but to avoid numerical problems we need to ensure that each MBS overlaps with at least one other MBS and avoid fine-tuning two overlaps to the exact same value.
 
\subsection{Conditions for parity blockade}
\label{subsec:BlockadeManifold}
Related research on Majorana box qubits connected to quantum dots \cite{Schulenborg_PRB2023} suggests that parity blockade from a mode $m$ connected to several MBSs $k$ with tunnel couplings $t_{mk}$ is established by fulfilling
\begin{equation}
\label{eq:parityCond}
 \sum_{k} t_{mk}^2 = 0.
\end{equation}
Note that in general, $t_{\alpha k} \in \mathbb{C}$ such that there exist nontrivial solutions of Eq.~(\ref{eq:parityCond}). Furthermore, the real and imaginary parts of Eq.~(\ref{eq:parityCond}) introduce two restrictions on the four-dimensional parameter space, introducing a two-dimensional parity blockade sub-manifold. In the following we will show how Eq.~\ref{eq:parityCond} is manifested in a transport measurement for both the Majorana box and Andreev box.

 We start by investigating the sub-manifold of tunnel couplings $t_0, t_1, t_2$ resulting in parity blockade. For a current to flow, at least one coupling needs to be non-zero. We use the gauge degree of freedom to choose $t_1 = t \in \mathbb{R}$. The remaining four variables are the absolute values $|t_l|$ and the phases $\phi_l$, $t_l = |t_l| e^{i \phi_l}$ for $l = 0, 2$. The MBS $\gamma_3$ connecting to the drain does not influence the parity blockade at the source, and we chose $t_3=t$.
 
\begin{figure}[t]
\includegraphics[width=\columnwidth]{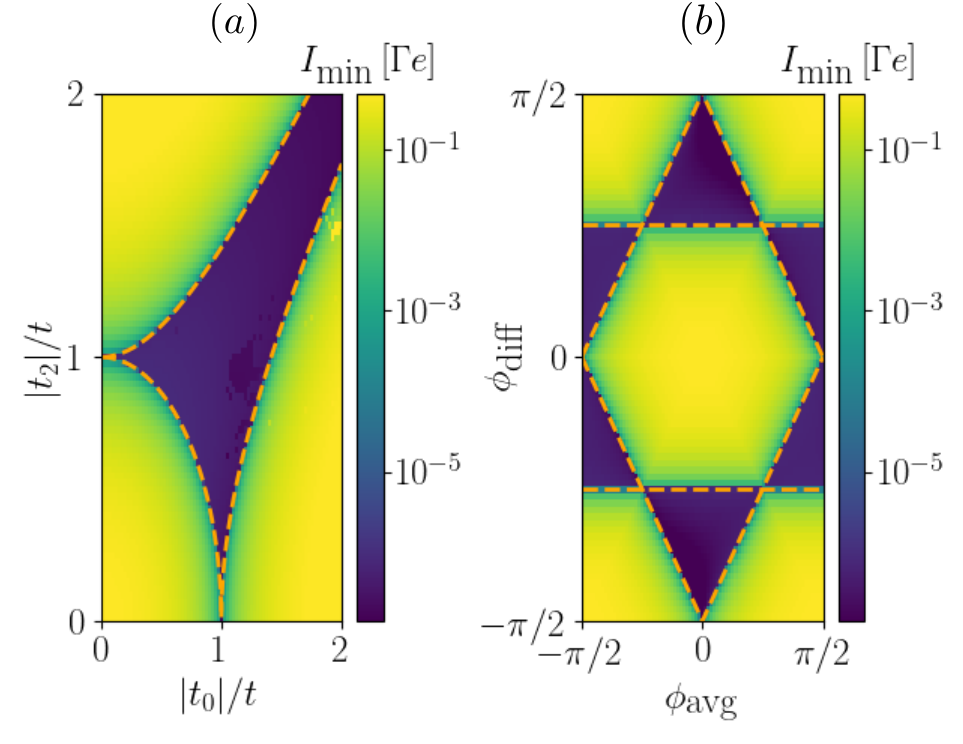}
\caption{Minimal current $I_\text{min}$ plotted on a logarithmic color scale for $(a)$ varying $|t_l|$ and optimizing $\phi_l$ and $(b)$ varying $\phi_\text{avg}, \phi_\text{diff}$ and optimizing $|t_l|$. The dark blue patches represent the parameter ranges with suppressed current due to parity blockade. In the remaining lighter regions parity blockade is not possible. With the orange dotted lines, we show the analytical results for the boundaries of the blockade regions found via Eq.~(\ref{eq:parityCond}). }
\label{fig:figure2}
\end{figure}

 Figure~\ref{fig:figure2} shows the regime where the current is suppressed due to the parity blockade. In Fig.~\ref{fig:figure2}$(a)$, we vary $|t_0|$ and $|t_2|$ and for each point ($|t_0|, |t_2|$) tune $\phi_0$ and $\phi_2$ to find the minimal possible current $I_\text{min}$. In Fig.~\ref{fig:figure2}$(b)$, we explore the opposite and vary $\phi_0$ and $\phi_2$ while tuning $|t_0|$ and $|t_2|$. For this, we introduce the average and difference of the phase
 \begin{align}
     \phi_\text{avg} = \frac{\phi_0 + \phi_2}{2}, && \phi_\text{diff} = \frac{\phi_0 - \phi_2}{2}.
 \end{align}
 
These results confirm that parity blockade defines a two-dimensional sub-manifold (dark, low-current regions in Fig.~\ref{fig:figure2}) within the four-dimensional parameter space.

We emphasize that Eq.~(\ref{eq:parityCond}) holds independent of whether the MBSs are topological or just trivial zero-energy ABSs expressed via Majorana (Hermitian) operators, see Fig.~\ref{fig:figure1}$(b)$. As discussed in Sec.~\ref{subsec:parityBlockade}, parity blockade in the Andreev box can be understood via a unitary rotation of the Majorana basis, reducing it effectively to the same mechanism that blocks the Majorana box. Accordingly, the system containing zero-energy ABSs mimics a Majorana box with additional MBSs within the wires decoupled from the lead. Consequently, the features of the parity/current blockade in Fig.~\ref{fig:figure2} remain qualitatively the same for the Andreev box. The only effect of the additional ABS parameters $\theta_{0, 1, 2, 3}$, introduced in Eq.~(\ref{eq:theta}), is to stretch/narrow Fig.~\ref{fig:figure2}$(a)$ along $t_0, t_1$ and to displace Fig.~\ref{fig:figure2}$(b)$ along $\phi_\text{avg}, \, \phi_\text{diff}$. We show an example of parity blockade in the Andreev box in Appendix \ref{sec:appendixParityBlockade}.

\subsection{Simple parity-blockade-based protocol to distinguish between MBSs and ABSs}
\label{subsec:currentDistinction}
Here, we present a protocol that yields different measurement results for the Majorana box compared to the Andreev box. During this protocol, we turn off certain connections between the source and the box. To avoid singular matrices $\hat{L}$ we define the minimum possible pinch off as $\Gamma_\text{min} = 10^{-6}$.

\begin{figure}[t]
\includegraphics[width=\columnwidth]{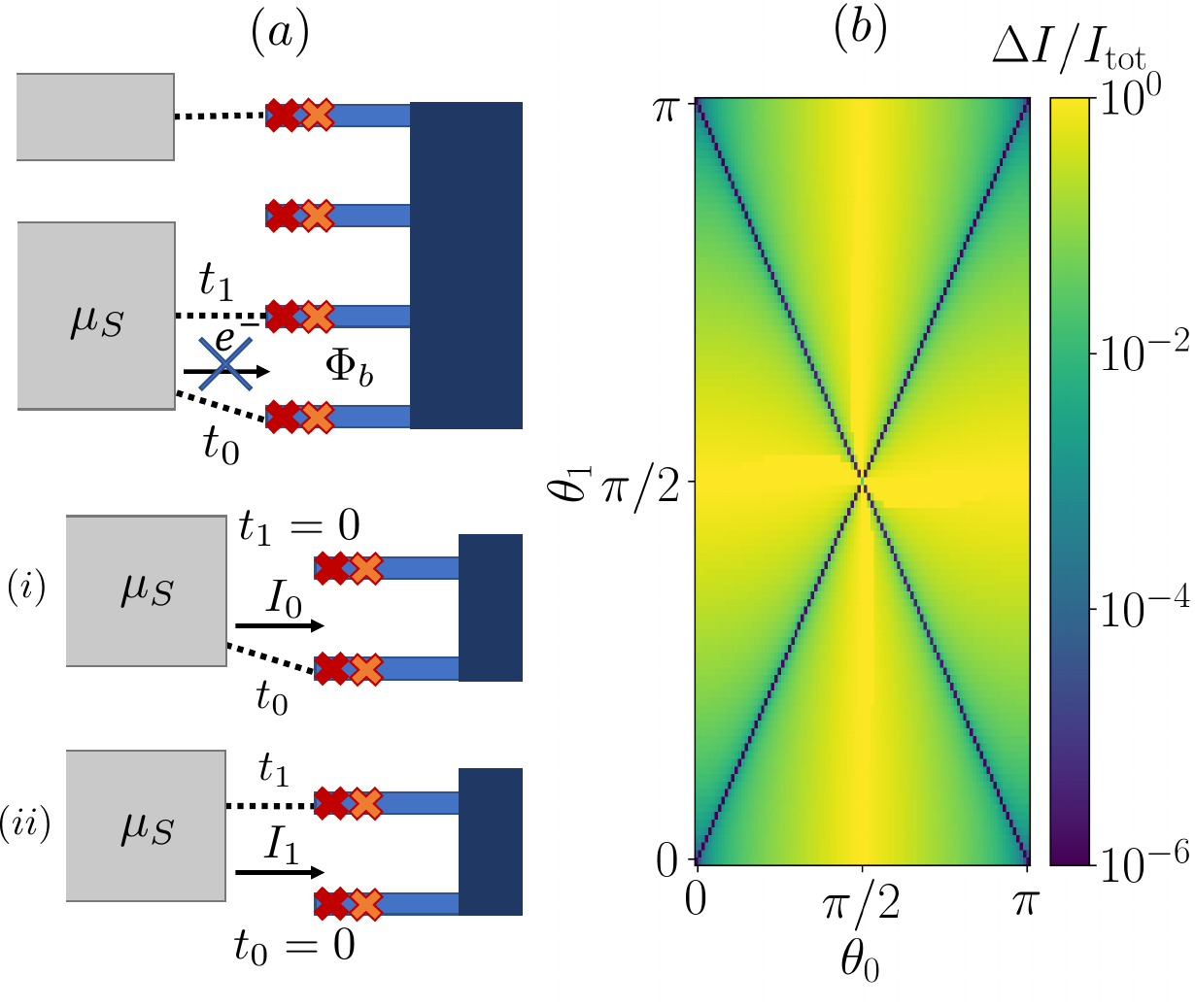}
\caption{Protocol to distinguish between ABSs and MBSs based on different currents after establishing and breaking parity blockade. $(a)$ We pinch off $t_2$ and establish parity blockade by tuning the loop $t_0, t_1$. Afterwards, we measure two currents, $I_0$ $(i)$ and $I_1$ $(ii)$, by pinching off $t_1$ or $t_0$. $(b)$ Normalized current difference $\Delta I/I_\text{tot}$ on a logarithmic scale for different values of $\theta_0, \theta_1$ in the ABSs. Compare to $\Delta I=0$ for the Majorana box.}
\label{fig:figure3}
\end{figure}

In this protocol, we will only need to couple the source to two wires and without loss of generality take $\mbox{$t_2=t_2'=0$}$ [see Fig.~\ref{fig:figure3}$(a)$], but other realization are possible as well \cite{Nitsch_PRB2022}.
The tunneling amplitudes $t_0, \, t_1$ and the flux $\Phi_{01}$ are tuned to establish parity blockade between the source and the box. Note that $\tilde{t}_0 = t_0$ and $\tilde{t}_1 = t_1$ in the case of the Andreev box due to our previous choice of the Majorana basis (Sec \ref{sec:ABSbox}). By establishing parity blockade we enforce a fixed relation between $t_0$ and $t_1$.

After the blockade is established, we keep the tunnel coupling to wire $l$ constant but pinch off the other one and measure the current $I_l \propto t_l^2$. We then repeat the same procedure but switch which tunnel coupling is pinched off. In the case of the Majorana box, parity blockade enforces that $t_0 = t_1$. Therefore, the current measurement yields $I_0 = I_1$. The Andreev box contains the additional degrees of freedom $\theta_0, \theta_1$. Except in the fine-tuned cases $\theta_1 = \theta_0$ or $\theta_1 = \pi - \theta_0$, parity blockade is reached for $|t_0| \neq |t_1|$ leading to different currents $I_0 \neq I_1$.

Figure~\ref{fig:figure3}$(b)$ shows the difference of the currents $\mbox{ $\Delta I = |I_1 - I_0|$ }$ normalized by the total current $ \mbox{$I_\text{tot} = I_1 + I_0$ }$ as a function of $\theta_0$ and $\theta_1$ for an Andreev box, measured according to the protocol described above. The value zero indicates a perfect imitation of the Majorana box. Exact zeros occur only on the diagonals $\theta_1 = \theta_0, \, \theta_1 = \pi - \theta_0$. The one-dimensional diagonals represent only a parameter space of volume zero within the two-dimensional parameter space. Therefore, only highly fine-tuned ABSs would yield the same results as MBSs. For an almost perfectly fine-tuned Andreev box, $\mbox{ $\theta_1 \approx \theta_0, \, \pi - \theta_0,$ }$ $\Delta I$ is finite but perhaps too small for detection.

\section{Fusion-rule protocol}
\label{sec:fusion}

Next, we introduce the fusion-rule protocol. We will start by introducing the time evolution of a Majorana box in the fusion protocol. Afterwards follows an analysis of the charge transfer during the protocol. We finish with a comparison to an Andreev box.

\subsection{Time evolution of the Majorana box}
\label{subsec:timeEvo}
\begin{figure}[t]
\includegraphics[width=\columnwidth]{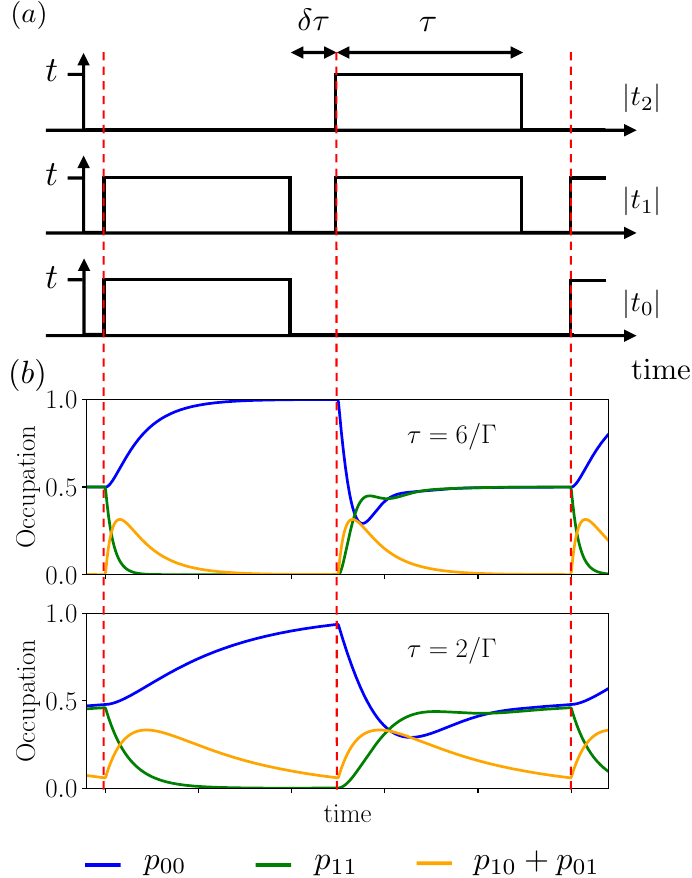}
\caption{$(a)$ Fusion protocol for a Majorana box. We establish two blockades with the connections $t_0, t_1$ and $t_1, t_2$, and repeatedly switch between them. The first pulse connects the box via $t_0, t_1$ establishing the z-blockade and the second connects the box via $t_1, t_2$ establishing the x-blockade. Each blockade is established for a waiting time $\tau$ with a changing time $\delta \tau$ in between the pulses. Throughout the whole protocol, the box is connected to the drain via $t_3$. $(b)$ Time evolution for the total even occupations $p_{00}, \, p_{11}$ and the sum of the total odd occupations $p_{10}+p_{10}$. The red dotted lines mark the connection in a new z/x-blockade to highlight the induced time evolution. We compare the results for a large waiting time $\tau = 6/\Gamma$ and a shorter waiting time $\tau = 2/\Gamma$. The changing time is fixed to $\delta \tau = \tau/4$.}
\label{fig:figure4}
\end{figure}

Figure~\ref{fig:figure4}$(a)$ sketches the fusion-rule protocol. 
Two different parity blockades are established via two different choices of the tunneling amplitudes $t_0, t_1, t_2$, where the phases are controlled by $\Phi_{01}$ and $\Phi_{12}$. We refer to the blockades as the z-blockade ($|t_0|=|t_1|=t, \, t_2=0$) and the x-blockade ($t_0= 0, \, |t_1|=|t_2|=t$). The names z-, and x-blockade are motivated by the orientation of the blocking states in terms of the basis $\ket{n_{01} n_{23}}$. They are
\begin{equation}
\begin{aligned}
    \text{z-blockade:} && & \hspace{0.cm} \ket{\psi_z} = \ket{00}, \\
    \text{x-blockade:} && & \ket{\psi_x} = \frac{\ket{00} + \ket{11}}{\sqrt{2}},
\end{aligned}
\end{equation}
which are eigenstates of the $\sigma_z, \sigma_x$ operators.

The protocol repeatedly switches between both blockades, thereby introducing two timescales, see Fig.~\ref{fig:figure4}$(a)$. First, $\tau$ constitutes the waiting time in a blocking configuration. Second, the changing time $\delta \tau$ represents the time between different blocking configurations. During $\delta \tau$ the system is completely decoupled from the source but still coupled to the drain. We assume that the ramping up and down of the tunnel amplitudes is much faster than the relevant timescales of the system, but sufficiently slow to avoid transitions to excited states. In the following, we will analyze the time evolution of the system during a full cycle, i.e., switching from z-blockade to x-blockade and then back to z-blockade.

To understand the system dynamics on an intuitive level we consider the limit of large waiting times, $\mbox{ $\tau \gg 1/\Gamma$ }$. The system is initialized in $\ket{\psi_x}$, then we switch on the z-blockade. Under z-blockade conditions, a charge cannot tunnel onto the box if it is in the state $\ket{00}$. Therefore, time evolution will eventually result in a projective measurement in the basis $\ket{00}$, $\ket{11}$, each occurring with $50 \, \%$ probability since the initial state is $\ket{\psi_x}$.
If the measurement yields $\ket{00}$, parity blockade prevents the electron from tunneling. But if the measurement results in $\ket{11}$, the charge can tunnel into the box, projecting the system onto $\ket{01}$. Afterwards, a charge tunnels out into the drain, projecting the box on the blocking state $\ket{\psi_z}=\ket{00}$. We can summarize the dynamics in the following sequence:
\begin{equation}
\label{eq:charge_cycle_blockade}
\begin{aligned}
    \ket{\psi_x} = \frac{\ket{00} + \ket{11}}{\sqrt{2}} \Rightarrow
    \begin{cases}
    \ket{00} \\[1.0ex]
    \ket{11} \rightarrow \ket{01} \rightarrow \ket{00}
\end{cases}.
\end{aligned}
\end{equation}
The time scale for each transition of the dynamics is $ \mbox{$\sim 1/\Gamma$}$. It is important to note that at each step in the dynamics, there is only one possible tunneling event due to the large charging energy of the box.

In total, either zero or one charge tunnels through the system. Both options happen with a $50 \, \%$ probability. One can check that the same holds for the opposite direction ($\ket{\psi_z} \rightarrow \ket{\psi_x}$), which finishes a full cycle of the protocol. Therefore, on average, the protocol transmits one charge per full cycle.

For a finite waiting time, it is not guaranteed that the system is fully projected onto $\ket{\psi_{x/z} }$ by the blockade. To achieve periodicity (equivalence between protocol cycles $n$ and $n+1$) we need to make sure, that the state at the end of one cycle is the same as in the beginning. Therefore, we run the protocol for 1000 cycles to ensure this self-consistency between states. Figure~\ref{fig:figure4}$(b)$ shows the time evolution of the system for a long waiting time $\tau=6/\Gamma$ (upper panel) and a short waiting time $\tau = 2/\Gamma$ (lower panel). It shows the time evolution during the 1001st protocol cycle. The self-consistency is seen in the equivalence of states at the first and last red dotted line.

We start our discussion with $\tau=6/\Gamma$, where the behavior follows the intuitive arguments above. Initially, the system is approximately in the state $\ket{\psi_x}$. We switch on the z-blockade, starting a time evolution into $\ket{\psi_z}$ intermediately occupying the states $\ket{10}, \ket{01}$. The moment we switch to the x-blockade, this causes the system to evolve from $\ket{\psi_z}$ to $\ket{\psi_x}$. Again it intermediately occupies $\ket{10}, \ket{01}$. The time evolution works qualitatively the same for the shorter waiting time $\tau = 2/\Gamma$. But in this case, the waiting time is too short to complete the transition between the blocking states. Note that the time evolution is also nontrivial during the changing time $\delta \tau$ because tunneling to the drain is still possible.

\subsection{Charge transfer in the Majorana box}

\begin{figure}[t]
\includegraphics[width=\columnwidth]{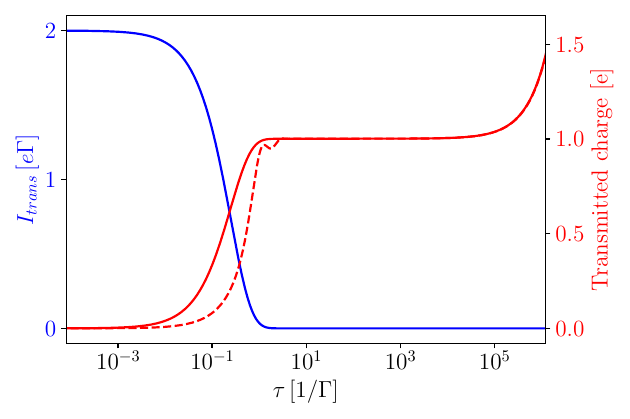}
\caption{Current $I_\text{trans}$ (solid blue line) and transmitted charge (solid red line) for the time evolution $\ket{\psi_z} \rightarrow \ket{\psi_x}$ as a function of the waiting time $\tau$ (logarithmic scale). The transmitted charge is multiplied by a factor of 2 to obtain the result of a full cycle of the fusion protocol. The dotted red line represents the transmitted charge during a full cycle for a self-consistent treatment of states in the fusion protocol. The changing time is fixed to instantaneous changes $\delta \tau = 0$.}
\label{fig:figure5}
\end{figure}

We now investigate how the waiting time affects the amount of transmitted charge. Figure~\ref{fig:figure5} shows the numerical result for the time evolution of the current through the source (solid blue line), which decays exponentially on the timescale $1/\Gamma$.
We obtain the transmitted charge (solid red line) by numerically integrating the current. The amount of transmitted charge develops a plateau at value $1 \, e$ for $\tau \gg 1/\Gamma$, confirming the intuitive reasoning in Sec.~\ref{subsec:timeEvo}. Although the current is suppressed by parity blockade, there exists a small, but finite remnant current $I_\text{rem}$ because of the MBS overlaps $\varepsilon$, $I_\text{rem} \propto \varepsilon^2/\Gamma$ \cite{Nitsch_PRB2022}. The contribution of this remnant current develops a magnitude on the order of $1 \, e$ if $\tau \gtrsim \Gamma/\varepsilon^2$ and leads to the upwards-bending of the charge plateau at long times. The same happens for a small deviation from the blockade conditions and we expect similar behavior also for a small but finite quasi-particle poisoning rate. 

Until now we investigated the charge transfer by initializing the system in a perfect projection on $\ket{\psi_z}$ and considered the time evolution in the x-blockade. For a self-consistent treatment of the blocking states (as described in Sec.~\ref{subsec:timeEvo}) the average amount of transmitted charge per cycle of the protocol is shown by the red dotted line in Fig.~\ref{fig:figure5}. We find a decrease in the transmitted charge already at larger $\tau$ compared to the charge transfer starting at a perfectly projected blocking state. For large waiting times $\tau \gg 1/\Gamma$ the previous results are recovered.

The experiment we envision aims at detecting the plateau at $1/\Gamma \ll \tau \ll \Gamma/\varepsilon^2$. The measured DC current is quantized to $I_\text{DC} = fe$, where $f=1/2(\tau + \delta \tau)$ is the frequency associated with a full cycle of the protocol.

\subsection{Fusion protocol result for the Andreev box}
\label{sec:ABSfusion}
Finally, we investigate the results of the fusion protocol for the Andreev box, demonstrating the absence of a quantized current. As explained in Sec.~\ref{subsec:BlockadeManifold}, also for ABSs we are guaranteed to find a setting for the tunnel couplings $|t_l|$ and magnetic fluxes $\Phi_{ll^\prime}$ to establish parity blockade in the previously introduced x- and z-blockade configurations. The blocking states are determined by the unitary operation $U^\dagger$ rotating the two ABSs into a basis where only one MBS is coupled from each of the two wires, see Fig.~\ref{fig:figure1}$(c)$. This rotation depends on the additional degree of freedom $\theta_l$ for each ABS. A further complication arises as the MBSs uncoupled from the lead, after application of the unitary $U^\dagger$, still have a small overlap with the coupled MBSs. These overlaps introduce dynamics on an additional time scale $1/\varepsilon$.
\begin{figure}[t]
\includegraphics[width=\columnwidth]{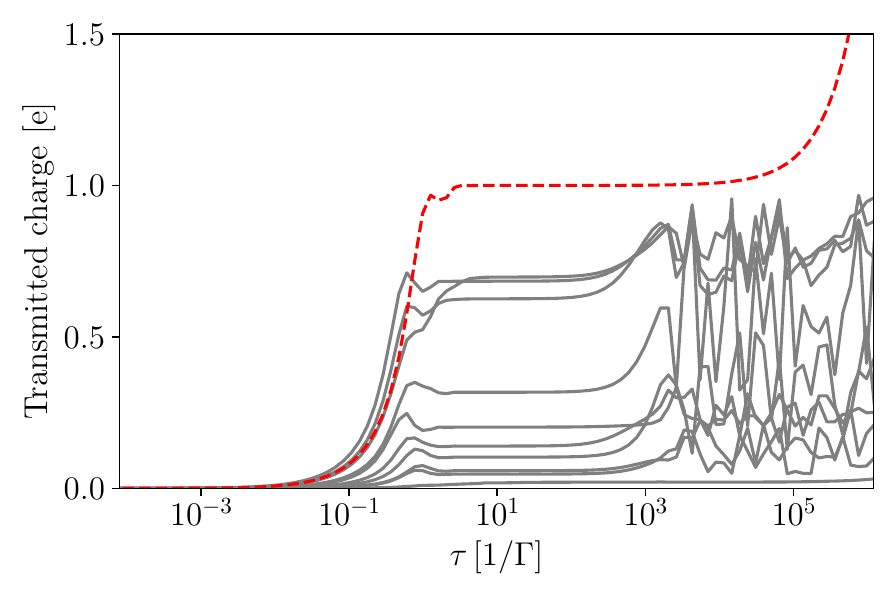}
\caption{Transmitted charge for the Andreev box in a full protocol cycle versus the waiting time $\tau$. The changing time is fixed to instantaneous changes $\delta \tau = 0$. Each gray line represents the results for one realization of randomly drawn ABS parameters $\theta_l \in [ 0, 2 \pi ] $ for each ABS on the sites $l=0, 1, 2, 3$. Depending on these parameters we determine the tunnel couplings and fluxes to establish parity blockade while connecting the source to ABSs $0, 1$ (z-blockade) and $1, 2$ (x-blockade). Afterwards, we run the fusion protocol as we did for the Majorana box, Fig.~\ref{fig:figure4}$(a)$. We run this protocol 1000 times to establish the same blocking states at each cycle. We compare the results to the previously obtained charge transfer for a Majorana box (red dashed line).}
\label{fig:figure6}
\end{figure}
Figure~\ref{fig:figure6} shows the results of the fusion protocol with an Andreev box.
The gray lines represent the results of the fusion protocol for ten configurations of $\theta_{0, 1, 2, 3}$, randomly drawn from a uniform distribution between $0$ and $2 \pi$, as a function of waiting time $\tau$. We also included the result for the pure Majorana box (red dashed line), for comparison. For short waiting times, $\tau < 1/\Gamma$, the transmitted charge for both systems tends to zero, as expected. In the other limit, $\tau > (\varepsilon^2/ \Gamma)^{-1}$, the transmitted charge increases linearly due to the remnant current $I_\text{rem} \propto \varepsilon^2/ \Gamma$ introduced by the finite overlaps of order $\varepsilon$.

The main difference between the Majorana box and the Andreev box appears for intermediate waiting times, $1/ \Gamma < \tau < (\varepsilon^2/\Gamma)^{-1}$. As discussed above, the Majorana box features a plateau at the predicted value of $1 \, e$. Detecting this quantized plateau for the Majorana box is the feature that allows us to distinguish it from the Andreev box, which shows a similar but non-quantized plateau before going over to oscillations at $\tau \approx \varepsilon^{-1}$. Therefore, we conclude that both the quantized current at the plateau and the stability of the plateau are signatures of MBSs which are very unlikely to appear in a similar system with nontopological zero-energy ABSs. 

Finally, we comment on why our fusion protocol is able to
distinguish between MBSs and zero-energy ABSs when some other fusion protocols fail to do so \cite{Clarke_PRB2017, tsintzis2023roadmap}. The difference lies in the number of involved MBSs. For an Andreev box, the number of fermionic states increases to four described by eight MBSs compared to just four MBSs in the Majorana box. These additional MBSs take part in the fusion process and add additional states to the possible fusion outcomes. This is not the case for the fusion protocols in, e.g., Refs. \cite{Clarke_PRB2017, tsintzis2023roadmap}, which considered trivial cases with only two fermionic states. 

\section{Conclusions}
In this paper, we have studied transport through a Majorana box using a QME, aiming to identify unique signatures of topological MBSs originating from the physics of parity blockade. Although parity blockade seems to be a special property of MBSs, we showed that nontopological ABSs also give rise to parity blockade that looks qualitatively similar in steady-state transport experiments. To distinguish between MBSs and ABSs, we first proposed a simple experiment based on comparing two current measurements with different configurations of lead tunnel couplings. Then we turned to a transport-based cyclic fusion protocol, where the system is interchangingly projected onto two different blocking states, also here by switching tunnel couplings on and off. For the Majorana box, we showed that the fusion rules result in a quantized DC current given by exactly the electron charge times the protocol frequency. We also discussed the limiting effects of MBS overlaps, quasiparticle poisoning, and deviations from the ideal blockade condition, showing that the current quantization can remain over a large frequency range. In contrast, for the Andreev box, we found a current that is not quantized and furthermore much less stable to changes in the protocol frequency. This not only provides a way to identify the presence of topological MBSs based on qualitative transport features, but also allows access to the fusion rules without the need for fast or single-shot readout. 

\begin{acknowledgments}
We acknowledge stimulating discussions with Jens Schulenborg and Athanasios Tsintzis and funding from the European Research Council (ERC) under the European Unions Horizon 2020 research and innovation programme under Grant Agreement No. 856526, the Spanish CM “Talento Program” (project No. 2022-T1/IND-24070), the Swedish Research Council under Grant Agreement No. 2020-03412, and NanoLund.
\end{acknowledgments}

\appendix

\section{Time-evolution first order von Neumann}
\label{sec:appendix1vN}
Here we briefly introduce the explicit form of the 1st order von Neumann master equation, for more details see \cite{Kirsanskas_CPC2017}. To obtain the dissipative part we define the tunneling rate matrix $\Gamma_{ba, a^\prime b^\prime}^r$
\begin{align}
    \Gamma_{ba, a^\prime b^\prime}^r = 2 \pi v_F T_{a \rightarrow b}^{r, \text{XB}} \, T_{b^\prime \rightarrow a^\prime}^{r, \text{XB}},
\end{align}
where the tunnel matrix elements are defined in Eq.~\ref{eq:T_r} for the Majorana box and in Eq.~\ref{eq:T_r_Andreev} for the Andreev box, and an integral over the Fermi distribution $f$
\begin{equation}
    2 \pi I_{ba}^{r\pm} = \int_{-K}^{K} \frac{f\left(\pm \frac{E-\mu_r}{T_r} \right)}{E-(E_b-E_a) +i \eta} \, dE,
\end{equation}
where $K$ is the bandwith and $\eta \rightarrow 0^+$. We obtain the time evolution of the density matrix as
\begin{equation}
\begin{aligned}
    i \partial_t \rho_{b b^\prime} =& (E_b - E_{b^\prime} ) \rho_{b b^\prime} \\
    &+ \sum_{b_{\prime \prime} \alpha} \rho_{b b^{\prime \prime} } \left[ \sum_a \Gamma_{b^{\prime \prime} a, a b^{\prime} } I_{ba}^{\alpha-} - \sum_c \Gamma_{b^{\prime \prime} c, c b^{\prime} } I_{cb}^{\alpha+ *} \right] \\
    &+ \sum_{b_{\prime \prime} \alpha} \rho_{b^{\prime \prime} b^\prime } \left[ \sum_c \Gamma_{b c, c b^{\prime \prime} } I_{c b^\prime}^{\alpha +} - \sum_a \Gamma_{ b a, a b^{\prime \prime} } I_{b^\prime a}^{\alpha- *} \right] \\
    &+ \sum_{a a^\prime \alpha} \rho_{a a^\prime} \Gamma_{b a, a^\prime b^{\prime} } [I_{b^{\prime} a }^{\alpha+*} - I_{b a^{\prime} }^{\alpha+}] \\
    &+ \sum_{c c^\prime \alpha} \rho_{c c^\prime} \Gamma_{b c, c^\prime b^{\prime} } [I_{c^{\prime} b }^{\alpha-*} - I_{c b^{\prime} }^{\alpha-}],
\end{aligned}
\end{equation}
where the indices $c$ and $a$ run over states with fixed electron number $N_c = N_b + 1$, $N_a = N_b -1$.

\section{Conditions for parity blockade in the Andreev box}
\label{sec:appendixParityBlockade}
We show that the Andreev box shows qualitatively the same regions for parity blockade as the Majorana box analyzed in Sec.~\ref{subsec:parityBlockade}.
\begin{figure}[h]
\includegraphics[width=\columnwidth]{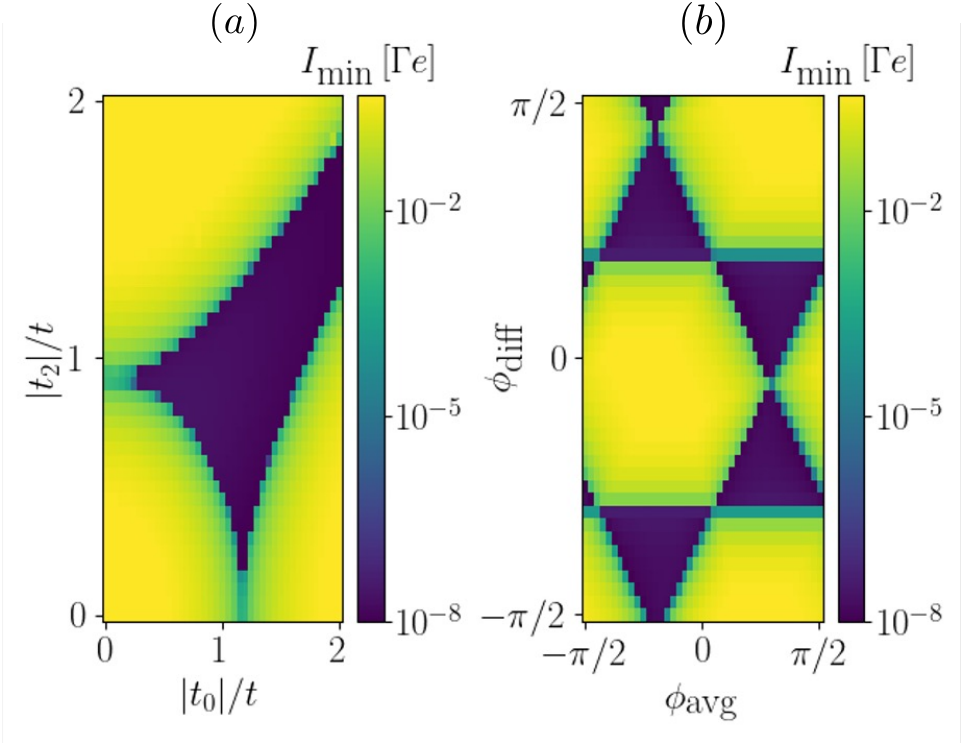}
\caption{Minimal current $I_\text{min}$ of an Andreev box plotted on a logarithmic color scale for $(a)$ varying $|t_l|$ and optimizing $\phi_l$ and $(b)$ varying $\phi_\text{avg}, \phi_\text{diff}$ and optimizing $|t_l|$. The dark blue patches represent the parameter ranges with suppressed current due to parity blockade. In the remaining lighter regions parity blockade is not possible. The Andreev parameters are chosen as $\theta_0=+0.3 \pi$, $\theta_1=-0.2 \pi$, $\theta_2=+0.1 \pi$, $\theta_3=-0.15 \pi$. The parity blockade regions are qualitatively the same as the parity blockade regions of the Majorana box.}
\label{fig:figure7}
\end{figure}
For a perfectly fine-tuned Andreev box Fig.~\ref{fig:figure7} would exactly coincide with Fig.~\ref{fig:figure2} for the Majorana box. Finite values for $\theta_{0, 1, 2}$ stretch/quench Fig.~\ref{fig:figure7}$(a)$ and shift Fig.~\ref{fig:figure7}$(b)$. Changing $\theta_3$ does not change Fig.~\ref{fig:figure7} as this site only connects to the drain and therefore has no influence on the parity blockade.

\bibliography{fusion}

\begin{thebibliography}{60}
\bibitem [{com()}]{comment1}%
  \BibitemOpen
  \href@noop {} {}\bibinfo {note} {Including the additional MBSs close to the
  trivial superconductor is equivalent to the $\theta \approx 0$ limit of the
  ABS model in Sec.~\ref{sec:ABSbox}, see also
  Sec.~\ref{sec:ABSfusion}}\BibitemShut {NoStop}%
\end{thebibliography}

\end{document}